\documentclass[11pt]{article}
\usepackage{fullpage}
\usepackage{amsfonts}
\usepackage{algorithmic}
\usepackage[boxed]{algorithm}
\usepackage{amsmath,amsthm,amssymb}
\usepackage{latexsym}
\usepackage{epic}
\usepackage{epsfig}
\usepackage{amscd}
\usepackage{url}
\usepackage{verbatim}
\usepackage{setspace}

\newtheorem{theorem}{Theorem}[section]

\newtheorem{claim}[theorem]{Claim}

\newcommand{\cross}{\times}

\newcommand{\set}[1]{\left\{ #1 \right\}}
\newcommand{\union}{\cup}

\renewcommand{\hat}{\widehat}

\def\Ex{\qopname\relax n{\mathbf{E}}}

\newcommand{\RR}{\mathbb{R}}
\newcommand{\RRp}{\RR_+}

\def\A{\mathcal{A}}

\def\D{\mathcal{D}}

\def\P{\mathcal{P}}
\def\R{\mathcal{R}}

\def\V{\mathcal{V}}

\newcommand{\eat}[1]{}

\newenvironment{lp*}{\begin{equation*}  \begin{array}{lll}}{\end{array}\end{equation*}}

\title{Mechanisms for Risk Averse Agents, Without Loss}

\author{
 Shaddin Dughmi\\
 Microsoft Research\\ 
{\tt shaddin@microsoft.com}
 \and
 Yuval Peres\\
Microsoft Research\\
 {\tt peres@microsoft.com}
}

\begin{document}

\maketitle

{
}

\begin{abstract}

Auctions in which agents' payoffs are random variables have received increased attention in recent years. In particular, recent work in algorithmic mechanism design has produced mechanisms employing internal randomization, partly in response to limitations on deterministic mechanisms imposed by computational complexity. For many of these mechanisms, which are often referred to as truthful-in-expectation,  incentive compatibility is contingent on the assumption that agents are risk-neutral. These mechanisms have been criticized on the grounds that this  assumption is too strong, because ``real'' agents are typically risk averse, and moreover their precise attitude towards risk is typically unknown a-priori. In response, researchers in algorithmic mechanism design have sought the design of universally-truthful mechanisms --- mechanisms for which incentive-compatibility makes no assumptions regarding agents' attitudes towards risk.

Starting with the observation that universal truthfulness is strictly stronger than incentive compatibility in the presence of risk aversion, we show that any truthful-in-expectation mechanism can be generically transformed into a mechanism that is incentive compatible even when agents are risk averse, without modifying the mechanism's allocation rule. The transformed mechanism does not require reporting of agents' risk profiles. Equivalently, our result can be stated as follows: Every (randomized) allocation rule that is implementable in dominant strategies when players are risk neutral is also implementable when players are endowed with an arbitrary and unknown concave utility function for money. 

Our result has two main implications: (1) A mechanism designer concerned with an objective  which depends only on the allocation rule of the mechanism can feel free to design a truthful-in-expectation mechanism, knowing that the risk-neutrality assumption can be removed by a generic black-box transformation.  (2) Studying universally-truthful mechanisms under the pretense of robustness to risk aversion is no longer justified.

\end{abstract}

\section{Introduction}

Auctions in which agents' payoffs are random variables have received increased attention in recent years. We are motivated by recent work in algorithmic mechanism design --- particularly in prior-free settings --- which has produced mechanisms employing internal randomization. The impetus for that work came from impossibility results for deterministic mechanisms: when constrained to polynomial time, randomized mechanisms provably yield superior approximation guarantees for objectives such as the social welfare than their deterministic counterparts, assuming standard complexity-theoretic conjectures such as $P \neq NP$. But randomization in agents' payoffs is nothing new; Bayesian mechanism design has always studied environments where players' types are drawn from a known prior, and consequently even a deterministic mechanism induces a random payoff for each agent. 

In many cases where  randomized auctions --- or auctions in randomized environments --- are studied, incentive compatibility is contingent on the assumption that agents are \emph{risk neutral} with respect to the random outcome resulting from the auction.  In prior-free settings, mechanisms that are incentive-compatible when agents are risk neutral have often been referred to as \emph{truthful in expectation}. Truthful-in-expectation mechanisms have been criticized on the grounds that the risk-neutrality assumption is too strong (e.g. \cite{Dobzin11,DNS06,DFK11,Vocking12}), sidelining these mechanisms as objects of mere theoretical interest. Consequently, random combinations of deterministic truthful mechanisms,  known as \emph{universally-truthful}, have received independent attention due to their robustness to agents' attitudes towards risk (e.g. \cite{DNS06,Vocking12}). This is despite the fact that the space of polynomial-time universally-truthful mechanisms is provably more restrictive, at least for the social welfare objective (see e.g. \cite{DD09}). Luckily, it is an easy observation that universal truthfulness is a strictly stronger requirement than robustness to risk aversion, at least when adopting the standard approach for modeling risk. This easy observation is the starting point for this note.


We show that, for any truthful-in-expectation mechanism, the assumption of risk neutrality can be removed without altering the mechanism's allocation rule. In other words, every (randomized) allocation rule that is implementable in dominant strategies when players are risk neutral is also implementable when players are endowed with an arbitrary unknown concave risk profile. A similar result holds for Bayesian settings, where randomness in an agent's payoff is a result of both internal randomization of the mechanism as well as randomness in other agents' types. Our result has two main implications: (1) A mechanism designer concerned with an objective  which depends only on the allocation rule of the mechanism can feel free to design a truthful-in-expectation mechanism, knowing that the risk-neutrality assumption can be removed by a generic black-box transformation.  (2) Studying universally-truthful mechanisms under the pretense of robustness to risk aversion is no longer justified. 

We note that the mathematical transformation underlying our result is similar to that used by  Es{\"o} and Futo \cite{EsoFuto99}, who consider a risk averse seller and risk neutral bidders. Nevertheless, we believe that the primary contribution of this note is not the underlying mathematics, but rather a framing much of the research in algorithmic mechanism design in as much as it relates to participants' attitudes towards risk.

\subsection{Results}

We show that any truthful-in-expectation mechanism can be converted to a mechanism that is dominant-strategy incentive compatible when players are risk averse, without modifying the mechanism's allocation rule. The intuition is as follows: a risk averse player facing a random monetary payoff $x$ can contract with a risk-neutral bank. The contract stipulates that the bank will always pay the player  $\Ex[x] - x$. This preserves the player's expected payoff, and removes all associated risk. Such a bank can be simulated by the principal running the auction mechanism, and we show that the simulation preserves incentive compatibility. In essence, the auction mechanism can insure players against risk by correlating monetary transfers with the realized payoffs.

An interpretation of our result is that any allocation rule implementable in dominant strategies when players are risk neutral is also implementable when players are risk averse. The transformation results in no loss in the principal's utility when the principal is risk neutral, and only increases agents' utilities. Moreover, the transformation does not depend on players' risk profiles, and does not require reporting of said profiles. While our results are motivated by randomized mechanisms in prior-free settings, we note that they immediately apply to Bayesian settings as well. Specifically, in a setting where player types are drawn from a known prior, the principal can remove all risk from the utilities of agents participating the auction by appropriately correlating the payments of an agent with the realized types of his competition in the auction.

One implication of this note is an  indictment of the notion of \emph{universal truthfulness} studied in algorithmic mechanism design. Specifically, mechanisms that are universally truthful are often cited as more desirable, due primarily to their robustness towards players' risk profiles, whereas truthful-in-expectation mechanisms are often thought of as constrained to quasilinear (risk neutral) utilities. This note demonstrates that, in fact, truthful-in-expectation mechanisms can be made incentive compatible for players with (unknown) risk averse utilities, at no loss in the objective so long as the objective is independent of the higher-order moments of agent payoffs and payments.

\subsection{Related Work}




Most of the literature in mechanism design assumes risk neutrality, both for the principal (also known as the seller) and the agents (the buyers). However, risk aversion has been studied, and we mention a sample of those works. Maskin and Riley \cite{MaskinRileyRisk} design a revenue-maximizing single-item auction in a Bayesian setting, where buyers are risk averse with given risk profiles.  Es{\"o} and Futo \cite{EsoFuto99} design  a single-item auction that is optimal for a risk-averse seller, in a Bayesian setting where buyers are  risk neutral. The mathematics underlying our result is similar to that underlying  the mechanism of \cite{EsoFuto99}; their mechanism is built from the revenue-maximizing auction of Myerson \cite{Myerson81}, and removes all risk from the seller's revenue by charging each player a random payment equal to that in \cite{Myerson81} in expectation, but appropriately correlated with the payments of other buyers such that the resulting revenue is deterministic.

More recent work has considered approximately optimal mechanisms for the seller's utility when the seller and/or buyers are risk averse. Sundararajan and Yan \cite{SY10} consider a multi-item Bayesian setting where the seller is risk averse, and buyers are risk neutral. They assume that the seller's risk profile is unknown, and achieve a  constant-factor approximation of the optimal seller utility simultaneously for all risk profiles. Bhalgat et al \cite{bhalgatrisk} consider settings with both risk averse buyers and sellers, with known risk profiles, and devise constant-factor approximations for the optimal seller utility.

\section{Preliminaries}

Let $[n]=\set{1,\ldots,n}$ denote a set of players. Let $\Omega$ denote a set of feasible outcomes. For each $i \in [n]$, let $\V_i$ denote a family of valuation functions for player $i$ mapping $\Omega$ to $\RR$, and let $\V=\V_1 \cross \V_2 \ldots \cross \V_n$. We consider direct revelation mechanisms of the form ($\A$,$p$) where $\A: \V \to \Omega$ is an \emph{allocation rule}, and $p: \V \to \RR^n$ is a \emph{payment rule}. We allow the mechanism to be randomized, in which case $\A(v)$ and $p(v)$ are random variables.

Such a mechanism is said to be \emph{truthful-in-expectation} if each player maximizes his expected \emph{payoff}, defined as as his value for the chosen allocation less his payment,  by reporting truthfully; precisely, if

\begin{equation}
  \label{eq:tie}
  \Ex[v_i(\A(v)) - p_i(v)] \geq \Ex[v_i(\A(v_{-i},v'_i)) - p_i(v_{-i},v'_i)]
\end{equation}
for all $i\in [n]$, $v \in \V$, $v'_i \in \V_i$, where the expectation is taken over the internal randomness of the mechanism $(\A,p)$. If a player $i$ is \emph{risk neutral}, i.e. his utility is simply his payoff, then reporting truthfully is a dominant strategy in such a mechanism. Truthful-in-expectation mechanisms can also be called \emph{dominant strategy incentive compatible for risk neutral players}.

We consider risk averse players where risk aversion is modeled as follows. A risk averse player $i$ with valuation $v_i$ is also equipped with a non-decreasing, concave function $u_i:\RR \to \RR$ denoting his utility for money. When faced with an random outcome $\omega \in \Omega$ and random payment $p_i$, the player's utility is defined as
\begin{equation}
  \label{eq:utility}
  u_i(v_i(\omega) - p_i)
\end{equation}

A mechanism $(\A,p)$ is said to be \emph{dominant strategy incentive compatible for risk averse players} if each player maximizes his expected utility by reporting truthfully; precisely, if

\begin{equation}
  \label{eq:tie_ra}
  \Ex[u_i(v_i(\A(v)) - p_i(v))] \geq \Ex[u_i(v_i(\A(v_{-i},v'_i)) - p_i(v_{-i},v'_i))]
\end{equation}
for all $i \in [n]$, $v \in \V$, $v'_i \in \V_i$, and non-decreasing and concave $u_i: \RR \to \RR$. When participating in such a mechanism, a risk averse player  maximizes his utility by bidding truthfully regardless of reports of other players.

\section{The Transformation}
\label{sec:main}
We now show that for every truthful-in-expectation mechanism $(\A,p)$, there is a mechanism $(\A,p')$ which is dominant-strategy incentive compatible for risk averse players. In other words, every allocation rule implementable for risk-neutral players is also implementable for risk-averse players. We stress that our mechanisms do not require reporting of the risk profiles, and the risk profiles may be arbitrary and unknown concave functions.

\begin{theorem}
  For every truthful in expectation mechanism $(\A,p)$, there is a payment rule $p'$, correlated with the outcome of $\A$, such that $(\A,p')$ is dominant-strategy incentive compatible for risk averse players. Moreover, $\Ex[p'_i(v)] = \Ex[ p_i(v)]$ for each player $i$ and reported valuation profile~$v$.
\end{theorem}

We define $p'$ as follows, correlated with the random outcome of the allocation rule $\A$
\begin{equation}
p'_i(v) = v_i(\A(v)) - \Ex[ v_i(\A(v)) - p_i(v)].
\end{equation}
Two claims are self-evident:
\begin{claim}\label{tie-still}
  $\Ex[p'_i(v)] = \Ex[p_i(v)]$ for each $i$ and $v \in \V$.
\end{claim}
\begin{claim}\label{det}
$v_i(\A(v)) - p'_i(v)$ is equal to its expectation, which is $\Ex[ v_i(\A(v)) - p_i(v)]$,  for \emph{every} realization of the mechanism's coins.  
\end{claim}

Claim \ref{tie-still} implies that $(\A,p')$ is also truthful-in-expectation. Claim \ref{det} implies that a player who bids truthfully faces a deterministic monetary payoff. Now, invoking Jensen's inequality completes the proof that $(\A,p')$ is incentive compatible for risk averse players. 

\begin{align}
   \Ex[u_i(v_i(\A(v)) - p'_i(v))] &= u_i(\Ex[ v_i(\A(v)) - p_i(v)])  \label{pf:1} \\
  &\geq u_i(\Ex[ v_i(\A(v_{-i},v'_i)) - p_i(v_{-i},v'_i))])  \label{pf:2} \\
&= u_i(\Ex[ v_i(\A(v_{-i},v'_i)) - p'_i(v_{-i},v'_i))]) \label{pf:3} \\
& \geq\Ex[u_i(v_i(\A(v_{-i},v'_i)) - p'_i(v_{-i},v'_i))] \label{pf:4}
\end{align}

Equality \eqref{pf:1} follows from claim \ref{det}, inequality \eqref{pf:2} follows from the assumption that $(\A,p)$ is truthful in expectation and from monotonicity of $u_i$, equality \eqref{pf:3} follows from claim \ref{tie-still}, and  \eqref{pf:4} follows from Jensen's inequality for concave functions.

\section{Additional Discussion}

The main takeaway from this note is the following: Once a mechanism designer has committed to an allocation rule, the risk neutrality assumption is \emph{without loss}. Specifically, the assumption of risk neutrality can be relaxed to risk aversion generically and without modifying the allocation rule. 

There are additional issues worth mentioning. First, while we posed our technical statements for prior free settings, a similar result holds in Bayesian environments. Second, this note weakens the most commonly invoked justification for the study of universally-truthful mechanisms, which have been a mainstay of algorithmic mechanism design for mainly historical reasons. Third, while our transformation can easily be implemented in polynomial time when an agent's expected payoff can be calculated efficiently, there are computational considerations in mechanisms where the expectation is unknown and can only be sampled. Fourth, there are valid critiques of our result, including the fact that transfers in the resulting mechanisms may flow in both directions --- from seller to buyer and vice versa --- even when only one direction is natural, and the fact that the transformation may be unjustified when the choice of allocation rule may itself be contingent on the risk neutrality assumption. We discuss these four issues below.

\subsection{Extension to Bayesian Settings}
In Bayesian mechanism design, the valuation profile $v$ is drawn from a prior distribution $\D$ which is common knowledge. In this context, incentive-compatibility is defined in reference to the prior $\D$. Assuming risk neutrality,  a mechanism $(\A,p)$ is \emph{Bayesian Incentive Compatible} if each agent $i$ with valuation $v_i$ maximizes his expected payoff by reporting truthfully, where the expectation is over the randomness is over draws of other agents' types, as well as the internal randomness of the mechanism. In other words, Bayesian incentive compatibility is equivalent to truth-telling being a Bayes-Nash equilibrium of the resulting game. 

As in the prior-free setting discussed in the rest of this note, risk neutrality can be relaxed to risk aversion in Bayesian settings. Specifically, observe that our transformation in Section \ref{sec:main} was agnostic to the source of randomness in the agent's payoff, and so was its proof of correctness. This implies the following analogous result.

\begin{theorem}
  For every Bayesian incentive compatible mechanism $(\A,p)$ in a setting with risk neutral players, there is a payment rule $p'$, correlated with the outcome of $\A$ and the types of the agents, such that $(\A,p')$ is Bayesian incentive compatible for risk averse players. Moreover, for each agent $i$ with fixed report $v_i$, we have that $\Ex[p'_i(v_i,v_{-i})] = \Ex[ p_i(v_i,v_{-i})]$, where the randomness is over the internal random coins of the payment rules as well as the types $v_{-i}$ of players other than~$i$.
\end{theorem}

\subsection{Implications for Universal Truthfulness}
Much of the literature on algorithmic mechanism design concerns \emph{universally-truthful} mechanisms. We posit three main reasons for this: (1) Due to technical barriers, impossibility results were constrained to universally-truthful mechanisms in the early days of the field (e.g. \cite{DN07a, PSS08}), (2) truthful-in-expectation mechanisms surpassing the guarantees of their universally-truthful counterparts were few and far between until recently, and (3) recently-discovered truthful-in-expectation mechanisms  appear impractical at first glance, particularly because they exploit the risk neutrality assumption to excess; for example by allocating the player many items with some probability, and nothing the rest of the time (e.g. \cite{DD09}). 

The study of universally-truthful mechanisms has been justified on the grounds of robustness to risk aversion (e.g. (e.g. \cite{Dobzin11,DNS06,Vocking12}),), an appealing argument in light of point (3) above. However, this justification misses an important distinction: universal truthfulness is a provably stronger assumption than robustness to risk aversion, specifically when risk aversion is modeled by a concave utility function for money. In fact, in many settings universally-truthful mechanisms are provably weaker, in terms of approximation the social welfare objective in polynomial time, than truthful-in-expectation mechanisms. Yet, as we show in this note, truthful-in-expectation mechanisms can be generically converted to mechanisms that are incentive-compatible for risk averse players without changing the allocation rule. This weakens the argument for the study of universally-truthful mechanisms, unless one is concerned with risk-seeking behavior.

\subsection{Computational Considerations}
It is easy to see that the only nontrivial requirement for efficient implementation of the transformation of Section \ref{sec:main} is that the expected payoff of each agent be calculable efficiently. This is indeed the case for some truthful-in-expectation mechanisms in the literature (e.g. \cite{DD09,LS05}). However, other recent examples do not admit a simple closed form for the expected payoff (\cite{DRY11,D11}). In such cases, it is not clear whether incentive compatibility in the presence of risk aversion  can be recovered, even in an approximate sense, in a generic way. However, we argue that multiplicative approximations of the expected payoff suffices for approximate incentive compatibility in the presence of risk aversion, and show an example where this is possible.

To make this discussion concrete, we examine the mechanism of Dughmi, Roughgarden and Yan~\cite{DRY11}  for combinatorial auctions  as an illustrative example. In combinatorial auctions,  the outcomes $\Omega$ are the set of partial allocations of items $[m]=\{1,\ldots,m\}$ among the players $1,\ldots,n$. We represent an allocation as a vector of disjoint bundles $(S_1,\ldots,S_n)$, where each $S_i$ is a subset of the items $[m]$. Each player $i$ is equipped with a valuation function $v_i:2^{[m]} \to \RRp$, and player $i$'s payoff from an allocation $(S_1,\ldots,S_n)$ is simply $v_i(S_i)$. One class of valuation functions tackled in  \cite{DRY11}, which we will focus on here, is \emph{succinct coverage valuations}. Specifically, $v_i$ is represented explicitly as a pair $(U^i, X^i)$, where $U^i$ is a set and $X^i$ is a family of $m$ subsets of $U^i$, indexed by the items $[m]$.   The valuation function $v_i$ is then defined as follows: $v_i(S) = |\union_{j \in S} X^i_j| $.\footnote{We note that, as defined here, elements of $U^i$ are unweighted. This is merely to simplify our presentation; our arguments apply equally well when each element of $U^i$ is equipped with a weight.}

The mechanism of \cite{DRY11} uses a randomized allocation rule $\A$, defined as follows. First, they let the set $\P$ be the family of fractional allocations; specifically, $\P = \set{x \in \RRp^{n \times m} : \forall j \in [m], \ \sum_{i=1}^n x_{ij} \leq 1 }$. Then, they associate each $x \in \P$ with a distribution $D_x$ over allocations; namely, the allocation assigning item $j$ to each player $i$ with probability $1-e^{-x_{ij}}$, independently for all items. The range $\R$ of the allocation rule is then defined as $\set{D_x: x \in \P}$. Fortuitously, for each profile $v$ of coverage valuations the expected social surplus of $D_x$ is a concave function of $x$. Therefore computing the $x^* = x^*(v) \in \P$ associated with the welfare-maximizing distribution $D_{x^*} \in \R$ is a convex optimization problem. Letting $\A(v)$ be a sample from $D_{x^*(v)}$, and charging each player a payment $p(v)$ with expectation equal to his  externality with respect to $\A(v)$,\footnote{An agent $i$'s externality with respect to an allocation rule $\A$ is $ \Ex[\sum_{j \neq i} v_j(\A(v_{-i},0)) - \sum_{j \neq i} v_j(\A(v))]$} gives a randomized member of the VCG family, and is therefore truthful in expectation.

Implementing this mechanism raises an important complication: $x^*$ can not, in general, be computed explicitly. In fact this is manifestly impossible, as $x^*$ may be irrational. This obstacle is overcome in \cite{DRY11} by exploiting the fact that $D_{x^*}$ need only be sampled, rather than written down explicitly. Therefore, by interleaving the sampling process with an approximation procedure for $x^*$ (using, e.g. the ellipsoid method), the distribution $D_{x^*}$ can be sampled in expected polynomial time as needed for implementing the mechanism.

Combining this idea with the reduction we propose in this note does not appear possible in general. Specifically, our reduction requires  explicit computation of each player's expected payoff from the mechanism, whereas the trick just described can merely sample it. Even resorting to approximation via random sampling does not appear to yield approximate incentive compatibility for risk averse players. Specifically, random sampling can obtain an additive $O(\epsilon v_i([m]))$-approximation for player $i$'s payoff with high probability. However,  due to concavity of player $i$'s utility, this does not lead to an $O(\epsilon u_i(v_i([m])))$ additive approximation for player $i$'s utility. 

Suppose instead that we are able to compute a \emph{multiplicative} $(1-\epsilon)$-approximation for player $i$'s payoff from the mechanism. Jensen's inequality now implies that this yields a $(1-\epsilon)$-approximation for player $i$'s utility. Plugging this into our reduction yields a $(1-\epsilon)$-approximately incentive compatible mechanism $(\A,\hat{p})$ for risk averse players, in the following sense:

\begin{equation}
  \label{eq:apxtie}
  \Ex[u_i(v_i(\A(v)) - \hat{p}_i(v))] \geq (1-\epsilon) \Ex[u_i(v_i(\A(v_{-i},v'_i)) - \hat{p}_i(v_{-i},v'_i))]
\end{equation}
Unfortunately, such multiplicative approximation of agent payoffs from a mechanism is not possible in general, and does not appear possible for the mechanism of \cite{DRY11}, specifically for players whose payoffs are exponentially smaller than their maximum possible value for an allocation. Nevertheless, in this case an approximately-truthful (in the multiplicative sense of \eqref{eq:apxtie}) modification of the mechanism of \cite{DRY11}, presented in \cite{DRVY11}, allows such an approximation and therefore yields an approximately-incentive compatible mechanism for risk averse players through our reduction. We omit the details.  We expect that, in most cases, such approximations will be possible.

\subsection{Critiques}


The main critique of our reduction, in our opinion, can equally be levied against many critiques of truthful-in-expectation mechanisms for optimizing social welfare. It is best described as a hypothetical. Suppose, for example, that a principal designs an allocation rule $\A$ that (approximately) maximizes the social welfare in the presence of risk neutral players, and a payment rule $p$ rendering $(\A,p)$ truthful in expectation. Our reduction yields $\hat{p}$ such that $(\A,\hat{p})$ is incentive compatible even when players are risk averse. However, in the presence of risk aversion,  $\A$ is no longer a desirable allocation rule because   \emph{it is optimizing the wrong objective!} Specifically, in the presence of risk averse bidders, the social welfare is no longer the sum of bidders' values for the allocation, but rather the sum of their utilities  plus the revenue of the principal (or the utility of the principal, when the principal is not risk neutral). 

In other words, the mere adoption of the sum of agent values as the objective implicitly assumes risk neutrality. Therefore, a truthful in expectation mechanism is already justified in light of this choice, and our transformation does not buy us anything unless we  concede that the objective being optimized is the wrong one. Nevertheless, even then our reduction does no harm, so long as we are unconcerned with the riskiness of  the principal's revenue. Moreover,  in situations where the choice of allocation rule is not influenced by the risk profiles of the agents, or is otherwise constrained to be independent of said profiles due to practical reasons or informational constraints, our reduction seems particularly appropriate.

Another deficiency of our reduction, discussed in the previous section,  is that it is not efficiently implementable in general, in particular for mechanisms not admitting a simple closed form for the distribution of their random allocation rule. Nevertheless, as described there, this can often be remedied by resorting to approximation. 

The final critique we mention concerns the direction of payments: since our payments are designed to insure an agent against risk, they may flow both from the principal to an agent and vice versa. This is true even in settings, such as auctions, where only one direction may be natural or practical.


{
\bibliography{agt}
\bibliographystyle{amsplain}
}

\end{document}